**Which are the best cities for psychology research worldwide?**

**A map visualizing city ratios of observed and expected numbers of highly-cited papers**

Lutz Bornmann [$], Loet Leydesdorff [§] and Günter Krampen [&]

[$] Max Planck Society, Hofgartenstr. 8, 80539 Munich, Germany; bornmann@gv.mpg.de.

[§] Amsterdam School of Communication Research (ASCoR), University of Amsterdam, Kloveniersburgwal 48, NL-1012 CX Amsterdam, The Netherlands; loet@leydesdorff.net.

[&] Leibniz-Institute for Psychology Information, University of Trier, Germany; krampen@uni-trier.de.

Corresponding author: Lutz Bornmann, bornmann@gv.mpg.de.

**Abstract**

We present scientometric results about world-wide centers of excellence in psychology. Based on Web of Science data, domain-specific excellence can be identified for cities where highly cited papers are published. Data refer to all psychology articles published in 2007 which are documented in the Social Science Citation Index and to their citation frequencies from 2007 to May 2011. Visualized are 214 cities with an article output of at least 50 in 2007. Statistical z tests are used for the evaluation of the degree to which an observed number of top-cited papers (top-10%) for a city differs from the number expected on the basis of randomness in the selection of papers.

Map visualizing city ratios on significant differences between observed and expected numbers of highly-cited papers point at excellence centers in cities at the East and West Coast of the United States as well as in Great Britain, Germany, the Netherlands, Ireland, Belgium, Sweden, Finland, Australia, and Taiwan. Furthermore, positive but non-significant differences in favor of high citation rates are documented for some cities in the United States, Great Britain, the Netherlands, the Scandinavian and the German-speaking countries, Belgium, France, Spain, Israel, South Korea, and China. Scientometric results show convincingly that highly-cited psychological research articles come from the Anglo-American countries and some of the non-English European countries in which the number of English-language publications has increased during the last decades.

# 1    Introduction

Since recently we have developed new approaches for the spatial visualization of concentrations of highly-cited papers using overlays to Google Maps. Bornmann, Leydesdorff, Walch-Solimena, and Ettl (in press) provided methods to map field-specific centers of excellence around the world using bibliometric data. These methods identify and agglomerate excellence in cities where highly-cited papers were published. Using colors and sizes for the marks, differences among cities in terms of performance rates can be visualized on maps. Bornmann and Waltman (in press) extended these methods and used a new approach with density maps for a spatial examination revealing regions of excellence. In contrast to Bornmann, Leydesdorff, Walch-Solimena, and Ettl (in press) who focus on cities, this approach is intended to visualize broader *regions* where highly-cited papers were published.

A further step in the development of these spatial visualization methods is described in Bornmann and Leydesdorff (in press). Our most recent methods do not only consider the quantitative numbers of highly-cited papers, but the observed number of highly-cited papers for a city is tested statistically against the expected number of highly-cited papers. For example, if authors located in a single city have published 1,000 papers, one would expect for statistical reasons that approximately 100 (that is, 10%) would also belong to the top-10% most-highly cited papers. An observed number of 70 highly-cited papers for this city may seem as a large number compared to other cities, but the specification of the expectation changes the appreciation. This approach has drawn considerable attention in science journalism (see, e.g., http://www.physorg.com/news/2011-03-european-team-scientific-relevance-city.html or

http://blogs.discovermagazine.com/80beats/2011/03/22/the-best-cambridge-london-and-worst-moscow-taipei-cities-for-science/).

The application of scientometric methods outside the physical and life sciences has sometimes been considered as questionnable (Moed, 2005). Although psychology does not belong to these "hard science" disciplines scientometric methods have increasingly been applied for evaluative purposes within the discipline (see, e.g., Endler, Rushton, & Roediger, 1978; Krampen, 2008; Nosek et al., 2010). A possible reason for this development is that publication and citation behavior in psychology converges at least in some of its subdisciplines (e.g., experimental psychology, bio- and neuropsychology, clinical psychology, social psychology) with the physical and life sciences which resulted during the last decades in increased publishing in English-language and internationally peer-reviewed journals (see, e.g., Krampen, Huckert, & Schui, in press; Krampen & Schui, 2011).

In this study, we provide a map for cities using all 2007 articles contained in the relevant so-called Subject Categories of the Social Science Citation Index (Web of Science, WoS, Thomson Reuters). Although Bornmann and Leydesdorff (in press) used psychology data as an example this study is more systematic and one of us as an expert validated the results. Bornmann and Leydesdorff (in press) based their psychology map only on a sample of all publications from one year; in this study all publications are included.

## 2 Methods

*Procedure to generate the underlying data*

The procedure to map the cities of the authors having published the top-10% most-highly-cited papers in a certain field is described in detail in Bornmann and Leydesdorff (in press). In the following, we give a short overview on the procedure.

The top-10% of papers with the highest citation counts in a publication set can be considered as highly cited (Bornmann, Mutz, Marx, Schier, & Daniel, 2011). In this study we follow this classification and focus on the top-10% of papers published in 2007 in psychology, using a citation window from 2007 up to the date of harvesting data from the WoS for this research (May 2011). All papers from 2007 having at least 16 citations each define the top-10% most cited psychology papers in this set. In a first step, all papers (n=21,528) with the document type "article" published in 2007 and belonging to the subject categories "psychology," "psychology, applied," "psychology, biological," "psychology, clinical," "psychology, developmental," "psychology, educational," "psychology, experimental," "psychology, mathematical," "psychology, multidisciplinary," "psychology, psychoanalysis," and "psychology, social" were downloaded from the WoS (Social Science Citation Index). We restricted the search to articles (as document types) since (1) the method proposed here is intended to identify excellence at the research front and (2) different document types have different expected citation rates, possibly resulting in non-comparable datasets.

After running in subsequent steps various programs which can be copied for free from http://www.leydesdorff.net/topcity/, the output file "ztest.txt" can be uploaded to the webpage of the GPS Visualizer (http://www.gpsvisualizer.com/map_input?form=data) to visualize the content. The file contains the city entries from the WoS data. If more than a single co-author but with an identical address is provided on a publication, this leads to a single city occurrence in the output. If the scientists are affiliated with departments in different cities, the different city names are used in the programs. The counting of occurrences in this study (so-called "integer counting") follows the procedure of how author addresses on publications are gathered by Thomson Reuters for inclusion in the WoS. In the output file "ztest.txt" the city entries from the WoS data are organized so that aggregated city occurrences can be visualized on a map, that is, provided with latitudes and longitudes (source of the coordinates: Google).

For the maps presented below we zoomed in on Europe and North America. Other regional foci can be studied using the full map at http://www.leydesdorff.net/psychology/cities.htm. One can inspect the ratio between observed and expected numbers of excellent papers for a specific city by clicking on the respective city node. Since the underlying data of a map from WoS (bibliographic data) and Google (geocoding coordinates; see here the comments on http://www.gpsvisualizer.com/geocoder/) are error-prone (Bornmann, et al., in press), we decided to visualize only cities (n=214) with an article output in 2007 of at least 50. There is a danger for cities in the data with a small number of papers that they result from private addresses of researchers or addresses of hospitals (but not from psychology departments). Furthermore, the use of a threshold of 50 provides us with a minimum of five papers expected in the top-10% and thus confirms with a requirement of using the statistical procedure described in the following.

For these 214 cities we could check the validity of the geocoding coordinates. In the case of (systemic) errors still on the maps, however, we appreciate and will respond to feedback.

*Statistical procedure*

The *z* test for two independent proportions (Sheskin, 2007, pp. 637-643) is used for evaluating the degree to which an observed number of top-cited papers for a city differs from the value that would be expected on the basis of randomness in the selection of papers (Bornmann & Leydesdorff, in press). *z* is positively signed if the observed number of top papers is larger than the expected number and negatively signed in the reverse case. An absolute value of *z* larger than 1.96 indicates statistical significance at the five percent level (p<.05) for the difference between observed and expected numbers of top-cited papers (marked with an asterisk *). In other words, the authors located at this city are outperformers with respect to scientific excellence in terms of this statistics. Due to the large number of city tests being conducted (n=214), especially highly significant p values (p<.01) should be considered as significant and interpreted (marked with at least two asterisks **; *** will indicate p<.001).

Using this statistical test, we designed the city circles which are visualized on the map using different colours and sizes. The radii of the circles are calculated by using:

| observed value – expected value | + 1

The "+1" must prevent the circles from disappearing if the observed ratio is equal to the expected one. Furthermore, the circles are coloured green if the observed values are larger than the expected values. We use dark green if *z* is statistically significant; light green indicates a positive, but statistically non-significant result. In the reverse case that the observed values are smaller than the expected values the circles are red or orange, respectively. They are red if the observed value is significantly smaller than the expected value and orange-red if the difference is statistically non-significant. If the expected value equals the observed value a city node is coloured grey.

## 3    Results

Figure 1 shows the location of authors in Europe having published highly-cited papers in psychology and the deviations of the observed from the expected number of top-cited papers per location (the circle radii) in 2007. If one clicks on a circle (at http://www.leydesdorff.net/psychology/cities.htm), a frame opens showing the number of observed versus expected values for the respective city, as well as an asterisk indicating whether the difference between the values is statistically significant or not. In Figure 1, for example, London is indicated by a very large dark green circle—the largest green circle in Europe— because of an observed value much larger than expected. In the description in a frame, the large and statistically highly significant difference between the observed ($n_o$=194) and the expected value ($n_e$=100.8) can be retrieved. Further large green circles on the map with a statistically highly significant difference ($p<.01$) between observed and expected values are visible for Oxford ($n_o$=40, $n_e$=20.0) and Cambridge ($n_o$=32, $n_e$=14.4) in Great Britain as well as Amsterdam ($n_o$=67, $n_e$=40.5). and Nijmegen ($n_o$=42, $n_e$=22.2) in the Netherlands.

Significant ($p<.05$) differences in favor of high citation counts are documented for the cities of Berlin ($n_o$=34, $n_e$=19.4), Bremen ($n_o$=13, $n_e$=5.3), Mannheim ($n_o$=15, $n_e$=6.6), Munster ($n_o$=14, $n_e$=5.6), and Leipzig ($n_o$=29, $n_e$=13.9) in Germany, Dublin in Ireland ($n_o$=18, $n_e$=6.8), Utrecht in the Netherlands ($n_o$=44, $n_e$=25.8), Louvain in Belgium ($n_o$=24, $n_e$=11.6), Stockholm in Sweden ($n_o$=38, $n_e$=20.9), and Oulu in Finland ($n_o$=13, $n_e$=5.0),

Furthermore, some more European cities are indicated with light green circles indicating a positive but statistical non-significant result. Most of these cities for which the relationship between expected and observed citations is positive, but not statistically significant are located in Great Britain (i.e., Aberdeen, Birmingham, Canterbury, Edinburgh, Lanark, Leeds, Manchester, Nottingham, Sheffield, and Southampton), the Netherlands (i.e., Groningen, Leiden, Rotterdam, and Tilburg), Germany (i.e., Cologne, Dusseldorf, Freiburg, Gottingen, Jena, and Munich), Belgium (Brussels, Ghent, and Maastricht), Switzerland (Bern, Geneva, Lausanne, and Zurich), Finland (Helsinki and Turku), Norway (Oslo and Bergen), Sweden (Gothenburg and Uppsala) and Spain (Madrid and Valencia). Single cities are located in Austria (Vienna), Denmark (Copenhagen) Finland (Helsinki), France (Paris), Italy (Milan), Norway (Bergen), Sweden (Stockholm), and it should be noted that only some of these cities are capitals.

The reverse case that the observed value is significantly smaller than the expected citation value is identified for no European city. However, there are many more cities which must be described as highly productive (i.e., psychology journal article output in 2007 of at least 50), but for which the relationship between expected and observed citations is negative without reaching statistical significance (see orange-red circles in Figure 1): Brighton, Bristol, Cardiff, Lancaster, Liverpool, and Southampton in Great Britain[1]; Frankfurt am Main, Heidelberg, and Tubingen in Germany; Padua and Rome in Italy; Barcelona and Granada in Spain; Marseille in France; Athens in Greece; Ankara in Turkey.

---

[1] An orange-red circle is also for Midlands, UK, on the map, an area comprising central England. Since the author addresses of the psychology articles did not contain further city indications, the circle refers to a broader area.

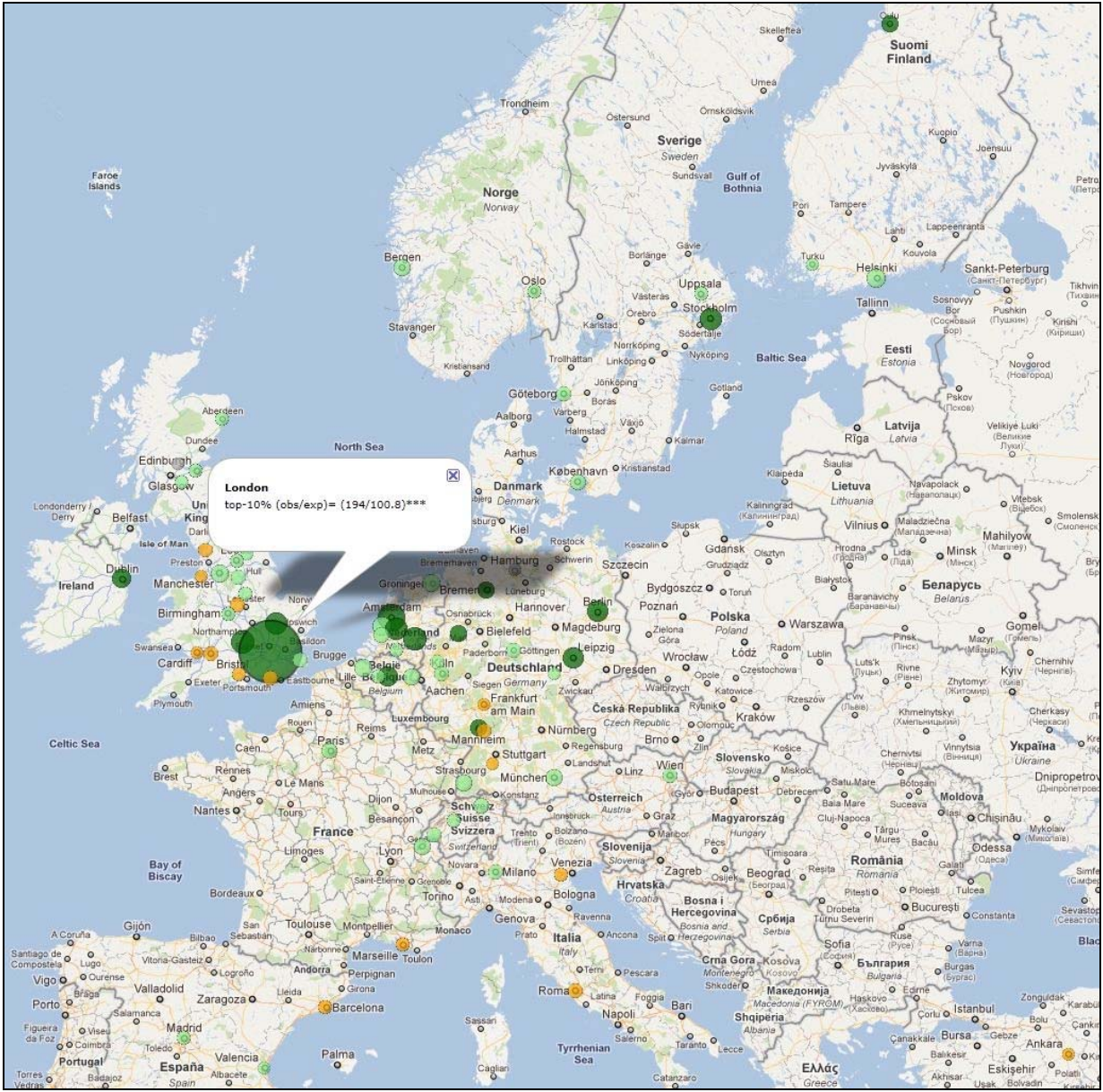

**Figure 1**. Cities in Europe with highly cited articles in psychology during 2007 (only cities are visualized with a total article output in 2007 of at least 50; see for the full map at http://www.leydesdorff.net/psychology/cities.htm)

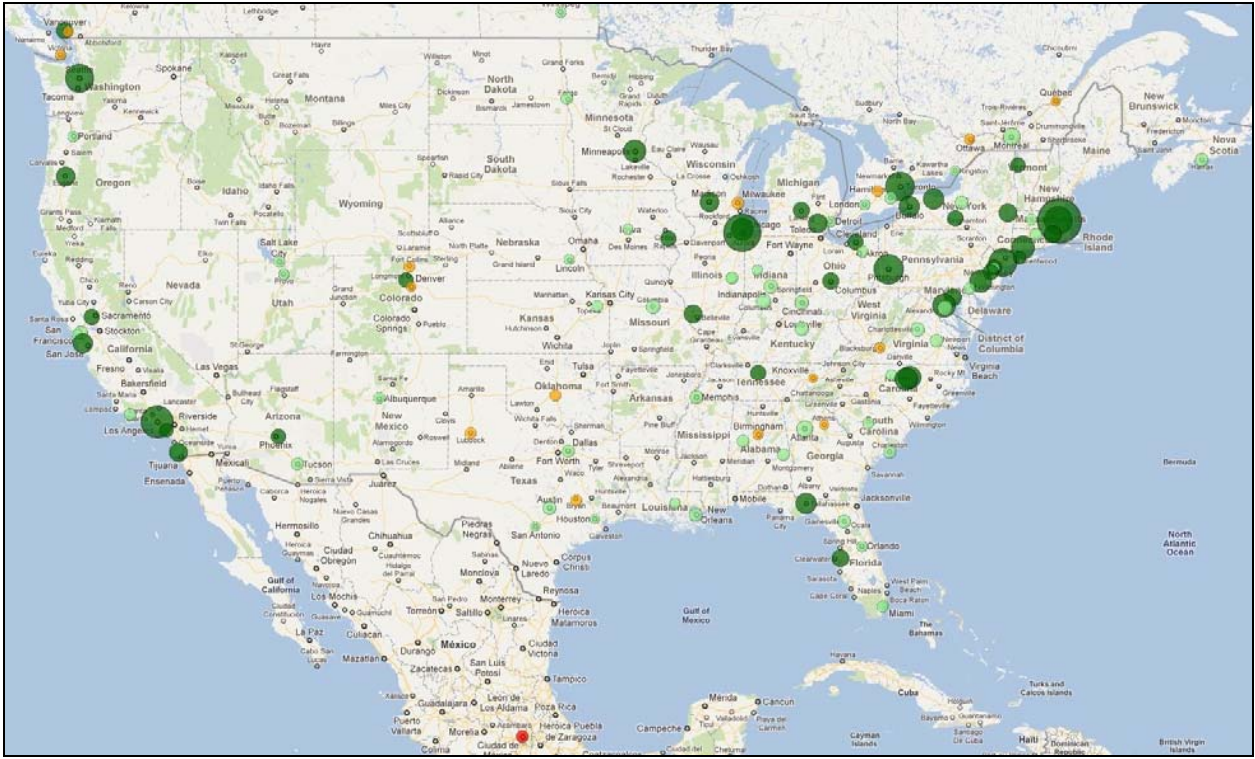

**Figure 2**. Cities in the USA with highly cited articles in psychology during 2007 (only cities are visualized with a total article output in 2007 of at least 50; see for the full map at http://www.leydesdorff.net/psychology/cities.htm)

Figure 2 shows the corresponding map focusing on the USA. There is a very high concentration of large green circles (indicating positive differences between observed and expected citation frequencies reaching statistical significance) at the East Coast which is completed by some more at the West cost as well as some in the North-East and South-East. The result that the best cities in 2007 for psychology research worldwide are located in the United States is supported further by the large number of light green circles, only a very few orange-red circles, and no red circles in Figure 2.

Extending the view to a global one, our results point to the fact that in addition to the above described cities in Europe and the United States only a few more cities show significantly positive differences of observed against expected citation frequencies of the psychology articles published in 2007: Sydney ($n_o$=45, $n_e$=27.8) and Parkville ($n_o$=19, $n_e$=7.8) in Australia as well as Kaohsiung in Taiwan ($n_o$=26, $n_e$=8.6). Furthermore, light green circles indicating a positive, but not statistically significant difference can be reported for Beer Sheva, Haifa, and Jerusalem (Israel), Adelaide and Melbourne (Australia), Christchurch and Dunedin (New Zealand), Singapore, Seoul (Taiwan) as well as Beijing (China).

Brisbane, Clayton and St. Lucia (Australia), Aichi[2] and Tokyo (Japan) as well as Hong Kong, Auckland (New Zealand), and Tel Aviv (Israel) are characterized by orange-red circles indicating a slight negative, but statistically non-significant difference for observed versus expected citation frequencies. Only Mexico City (Mexico), Sao Paulo (Brasilia), and Taipei (Taiwan) show statistically significant negative differences for observed versus expected citation numbers (red circles on the map) indicating an underrepresentation in the gain of top-10% citations.

## 4 Discussion

The maps presented in this paper allow for an analysis revealing centers of excellence in psychology around the world using scientometric data. Based on WoS data, field-specific excellence can be identified in cities where highly-cited papers were published. Compared to the mapping approaches published hitherto (see the Introduction), our approach is analytically oriented by allowing the assessment of an observed number of excellent papers for a city against the expected number. With this feature, this approach can not only identify the top performers in output but the "true jewels" in psychology. These are cities in which authors are located who publish significantly more top-cited papers than can be statistically expected. Since these cities do not necessarily have a high output of highly-cited papers, our approach normalizes for size.

---

[2] Aichi is a prefecture of Japan. Most of the author addresses in Aichi belong to Nagoya, the capital of Aichi.

For the example of psychology, the results presented show impressively that in our times most research results are published and cited within the Anglo-American countries: Far most of the best cities for psychology research worldwide are located in the United States and Great Britain. This is completed only by some cities in the Netherlands, Israel, Australia, the Scandinavian, and the German-speaking countries.

In addtion to research resources in terms of personnel and finance in the United States, it can be assumed that the recent incentive to publish research results in the English language will contribute to this result pattern. At the very latest, in the second part of the 20$^{th}$ century, English has become the language of the sciences. Anglicization of the sciences occurred rapidly in the natural sciences, particularly in areas of discovering facts and natural laws, which are not or—at least—only weakly and indirectly dependent on culture and socialization, i.e., language, nationality, politics, etc. The Anglicization of the arts and humanities occurs more gradually and less extensively because of its direct dependency and relationships to culture and socialization, e.g., national and cultural specifics of educational systems, social norms, traditions, etc. Somewhat in between the faster moving natural sciences and the slower arts and humanities are the social sciences and, especially, psychology, because its research topics (i.e., behavior and experience or—more contemporary—action and cognition) must be analyzed from both a natural science and a humanities perspective (see, e.g., Krampen, et al., in press).

For example, the psychobiological and neuropsychological determinants and correlates of behavior and experience are under psychological study as well as the social, cultural and psychodynamic aspects of action and cognition. Thus, psychological research can incorporate the natural science methodology as well as the (often more qualitative) methodology of the arts and humanities which includes the methods (e.g., statistics) of the social sciences. For the sake of optimizing the international visibility, reception, and impact of the results of psychological research in the non-English-speaking countries, fierce discussions in the research communities took and take place. A conclusion was that researchers should attempt to publish their empirical findings more frequently in Anglo-American and English-language journals and less frequently—or not at all—in national, non-English journals. In the recent past and to date, similar discussions are taking place in many larger non-English psychology research communities, such as France, Germany, Italy, and Spain. Some other, however smaller European research communities –Scandinavia and the Netherlands—began contributing to modern psychology increasingly in the English language after World War II.

Despite the advantages of our approach by mapping observed versus expected numbers, we recognize the limitations inherent to these bibliometric exercises (see here Bornmann & Leydesdorff, in press): (1) Publications are among several types of scientific activities. (2) It is not always the case that the addresses listed on the publication reflect the locations where the reported research was conducted. (3) Our method does not consider the different positions of authors on a paper. A paper is counted for a city if at least one co-author is located there. (4) Citation counts (and thus the categorization of publications as highly cited) are a function of many variables in addition to scientific quality (Bornmann & Daniel, 2008). Last, but not least, it should be noted that we present a spotlight on all psychology papers published in 2007 which are documented in the Social Science Citation Index covering many, but not all psychology journals while maintaining its focus on English-language publications.